\documentclass[reprint, aps, prd, preprintnumbers, amsmath, amssymb, superscriptaddress, nofootinbib]{revtex4-1}
\pdfoutput=1

\usepackage[utf8]{inputenc}
\usepackage{amssymb,amsmath,bbold,slashed}
\usepackage{graphicx,soul,caption}
\usepackage{booktabs,multirow}
\usepackage[usenames,dvipsnames]{xcolor}
\usepackage[unicode=true, pdfusetitle, 
bookmarks=true, 
bookmarksnumbered=false, 
bookmarksopen=false, 
citecolor=blue, 
breaklinks=false, 
pdfborder={0 0 1}, 
backref=false, 
colorlinks=true
]{hyperref}

\newcommand{\beq}{\begin{equation}}
\newcommand{\eeq}{\end{equation}}
\newcommand{\bea}{\begin{eqnarray}}
\newcommand{\eea}{\end{eqnarray}}

\newcommand{\nn}{\nonumber}

\newcommand{\keV}{{\rm keV}}
\newcommand{\GeV}{{\rm GeV}}

\newcommand{\Br}[1]{{\rm BR}\left({#1}\right)}

\newcommand{\Ug}{\rm U(1)_\phi}
\newcommand{\e}{{\rm e}}

\begin{document}
\setlength{\parindent}{0pt}
\preprint{FTUV-20-0717, IFIC/20-36}


\title{An anomaly-free leptophilic axion-like particle and its flavour violating tests}

\author{C. Han}
\email[]{hanchch@mail.sysu.edu.cn}
\affiliation{School of Physics, Sun Yat-Sen University, Guangzhou 510275, China}

\author{M.L. López-Ibáñez}
\email[]{maloi2@uv.es}
\affiliation{CAS Key Laboratory of Theoretical Physics, Institute of Theoretical Physics 
 	    \\ Chinese Academy of Sciences, Beijing 100190, China.}
\author{A. Melis}
\email[]{aurora.melis@uv.es}
\affiliation{Departament de Física Tèorica, Universitat de València \& IFIC, Universitat
 	    de València \& CSIC, \\ Dr. Moliner 50, E-46100 Burjassot (València), Spain}
\author{O. Vives}
\email[]{oscar.vives@uv.es}
\affiliation{Departament de Física Tèorica, Universitat de València \& IFIC, Universitat
 	    de València \& CSIC, \\ Dr. Moliner 50, E-46100 Burjassot (València), Spain}	    
\author{J.M. Yang}
\email[]{jmyang@itp.ac.cn}
\affiliation{CAS Key Laboratory of Theoretical Physics, Institute of Theoretical Physics 
 	    \\ Chinese Academy of Sciences, Beijing 100190, China.}
\affiliation{School of Physical Sciences, University of Chinese Academy of Sciences, Beijing 100049, China
 	    } 	    

\begin{abstract}
\noindent
Motivated by the recent Xenon1T result, we study a leptophilic flavour-dependent anomaly-free axion-like particle (ALP) and its effects on charged-lepton flavour violation (CLFV).
We present two representative models. 
The first one considers that the ALP origins from the {\it flavon} that generates the charged-lepton masses.
The second model assumes a larger flavour symmetry such that more general mixings in the charged-lepton are possible, while maintaining flavour-dependent ALP couplings.
We find that a keV ALP explaining the Xenon1T result is still viable for lepton flavour violation and stellar cooling astrophysical limits.
On the other hand, if the Xenon1T result is confirmed, 
future CLFV measurements can be complementary to probe such a possibility.
\end{abstract}

\maketitle

\section{Introduction}
Recently, the Xenon collaboration reported the observation of an excess in the electron recoiling energy around the keV scale in the Xenon1T detector \cite{Aprile:2020tmw}.
Shortly after its announcement, a lot of theoretical work has been done to interpret the results in the context of axion-like particles (ALPs) \cite{Takahashi:2020bpq, Bloch:2020uzh, Dent:2020jhf, Cacciapaglia:2020kbf, Li:2020naa, Long:2020uyf, Athron:2020maw, Croon:2020ehi, Croon:2020oga}, 
dark matter \cite{Kannike:2020agf, Alonso-Alvarez:2020cdv, Fornal:2020npv, Harigaya:2020ckz, Su:2020zny, Du:2020ybt, Bell:2020bes, Chen:2020gcl, Dey:2020sai, Choi:2020udy, Paz:2020pbc, Cao:2020bwd, Primulando:2020rdk, Nakayama:2020ikz, Jho:2020sku, Bramante:2020zos, Baryakhtar:2020rwy, An:2020bxd, Zu:2020idx, Zioutas:2020cul, An:2020tcg, DelleRose:2020pbh, Arcadi:2020zni, Choudhury:2020xui, Choi:2020kch}, neutrinos \cite{Shoemaker:2020kji, Boehm:2020ltd, Bally:2020yid, AristizabalSierra:2020edu, Buch:2020mrg, Khan:2020vaf, Ge:2020jfn, Chiang:2020hgb, He:2020wjs, Babu:2020ivd} and solar axions \cite{Gao:2020wer, Budnik:2020nwz, Coloma:2020voz}, which, however, are subject to stringent constraints from stellar cooling \cite{DiLuzio:2020jjp, DeRocco:2020xdt, Dessert:2020vxy}.
In this work we focus on the ALP framework.
This possibility assumes the existence of an ALP with a mass of a few keV and a relatively weak coupling to the electron.
However, constraints from X-ray observations forbid the existence of an anomalous coupling of the ALP to photons for $m_a \gtrsim 0.1$ keV.
An anomaly-free ALP with respect to $U(1)_{\rm em}$ can avoid these bounds.
With the SM fermionic particle content (plus right-handed neutrinos), only hypercharge and $B-L$ are completely anomaly-free with family universal charges. Hypercharge can be immediately discarded as it can not be broken above the electroweak scale. The breaking of $B-L$-gives rise to a pseudo-Goldstone boson, the Majoron, coupling only to neutrinos at tree-level. Although it couples to charged leptons at one-loop level, the $B-L$ breaking scale is related to the right-handed Majorana masses that is necessarily too high to explain Xenon1T result with our minimal particle content. Lepton number, also anomaly free with respect to  electromagnetism, faces the same problem with family universal charges.
Thus, we have to consider a U(1) symmetry with family-dependent charges which, as we will see, necessarily implies flavour-changing couplings between the ALP and the SM-lepton sector. 
If the excess is confirmed in the future, it will be necessary to investigate the lepton flavour violating signatures of this particle in low-energy experiments.
In this paper, we consider the flavour violation effects induced by such anomaly-free ALP, and 
we show that LFV measurements are essential to probe this possibility. 

Our paper is organized as follows: in section \ref{sec:Models}, we present two different models, both flavour dependent, but with distinct mixing patterns; the most important constraints to our models are collected in section \ref{sec:Constraints}; in section \ref{sec:results}, we present our results and discuss how these flavoured models can be tested by LFV data; finally, section \ref{sec:conclusions} is dedicated to our conclusions.
\section{Models}
\label{sec:Models}
We consider a $\Ug$ global symmetry spontaneously broken by the vev of a complex scalar field, $\phi$, whose angular component is identified with an ALP.
We propose two models with flavour dependence on the lepton sector and evaluate the importance of present and future experiments on lepton flavour violating (LFV) decays.
In the first model, the presence of the ALP is directly connected to the SM flavour puzzle and the breaking of the $\Ug$ is the only responsible of the observed hierarchy among the lepton generations. 
Instead, Model II generalizes the previous structure assuming the existence of a larger symmetry, which includes $\Ug$, whose breaking produces the Yukawa structures at high energies. In this way, we can partly decouple the non-anomalous flavour-dependent $\Ug$ charges from the observed leptonic masses and mixings. 
In both models, below the $\Ug$-breaking scale, the ALP has flavour-dependent couplings.

\subsection{Model I: hierarchical mixing}
\label{sec:ModelI}
Flavour symmetries à la Froggatt-Nielsen \cite{Froggatt:1978nt} offer an attractive solution to the origin of the observed hierarchy among the charged-fermion families.
In its simplest version, the spontaneous breaking of a U(1) flavour symmetry by the vev of a scalar field, usually called {\it flavon}, generates it as powers of the ratio between its vev, $v_\phi$, and $\Lambda$, the scale at which the heavy fields mediating the processes live, $\epsilon=|v_\phi/\Lambda|$.\\
In Model I, we identify this symmetry with the global $\Ug$ 
so that the angular component of the {\it flavon} corresponds to the ALP.
The case of the anomalous QCD axion has been previously explored in \cite{Ema:2016ops, Calibbi:2016hwq}, with the scalar receiving the name of {\it flaxion} or {\it axiflavon}.\\
As usual, in flavour models, distinct mixing patterns can be derived for different charge assignments.
Here we focus on the leptonic sector, hence quarks are assumed to be uncharged under the symmetry.
Besides, a sufficient condition to obtain an electromagnetic anomaly-free ALP is:
\bea
   \sum_i Q_{L_i}=0\quad, \quad\sum_i Q_{e_i}=0\,.
\eea
Then, we choose the charges under $\Ug$ of the left-handed leptons as $L(1,0,-1)$ and those of right-handed leptons to be $e(-1,0,1)$.
Such charge assignment is  crucial to generate the Froggatt-Nielsen structure.
Two Higgs doublets are introduced with charge 0 and -2, and an additional $Z_2$ symmetry is imposed as in the type-X 2HDM \cite{Chun:2018vsn} so that the only odd fields are $H_2\to -H_2$, $e \to -e$ and $N_R \to -N_R$.
The Higgses, $H_1$ and $H_2$, only couple to quarks and leptons, respectively. 
To summarize, the following particles and charges under $\Ug\times Z_2$ are considered for Model I:
\bea
    & H_1(0;\, 1),~ H_2(2;\, -1),~ \phi(1;\, 1), & \nn \\
    & L(1, 0, -1;\, 1),\,  e(-1,0,1;\, -1),\, N_R(0,0,0;\, -1)\,. & \hspace{1cm} \label{ModI-charges}
\eea
From \eqref{ModI-charges}, it can be seen that the anomalies cancel for both the left- and right-handed sector. 
The most general scalar potential is
\begin{eqnarray} \label{eq:VH}
V(H_1, H_2, \phi) &&= m_1^2 H_1^\dagger H_1+ m_2^2 H_2^\dagger H_2 +\lambda_1 (H_1^\dagger H_1)^2   \nonumber \\
&&  + \lambda_2 (H_2^\dagger H_2)^2 + \lambda_3 (H_1^\dagger H_1) (H_2^\dagger H_2)\\
&& +\lambda_4 |H_1 \cdot H_2|^2 + m^2  H_1 \cdot H_2  + \lambda (\phi^\dagger \phi- v_\phi^2 )^2, \nonumber
\end{eqnarray}
where we also add a soft breaking term $m^2 H_1 \cdot H_2$ for the $\Ug\times Z_2$ symmetry, then the ALP gets a mass around $m^2/v_\phi$.
The corresponding Yukawa terms are:
\begin{eqnarray}
\label{eq:LY}
\mathcal{L}_{Y} & \supset & Y_u\, \bar Q \widetilde H_1 u \;+\; Y_d\, \bar Q H_1 d \;+\; c^e_{ij}\, \epsilon^{n^e_{ij}}\, \bar L_i\, \widetilde H_2\, e_j \nonumber \\
& & +\; c^\nu_{ij}\, \epsilon^{n_{ij}^\nu}\, \bar L_i\,  H_2\, N_j \;+\; \left(M_R\right)_{ij}\, N_{R_i}\, N_{R_j}^c,
\end{eqnarray}
with $c^e_{ij}$ and $c^\nu_{ij}$ ${\cal O}(1)$ coefficients and {\small $n_{ij}^e=q_{L_i}-q_{e_j}+q_{H_2}$, $n^\nu_{ij}=q_{L_i}-q_{N_{R,j}}-q_{H_2}$}.
In Model I, we have
\begin{equation}
\label{eq:CargemModIa}
    n_{ij}^e ~=~ \begin{pmatrix}
        4  & 3  & 2\\
        3  & 2  & 1 \\
        2  & 1  & 0 \\
        \end{pmatrix},                              
    \hspace{1.cm}
    n_{ij}^\nu ~=~ \begin{pmatrix}
        1  & 1  & 1\\
        2  & 2  & 2 \\
        3  & 3  & 3 \\
        \end{pmatrix}.
\end{equation}
Once the EW symmetry is broken by the Higgs vev, $v_{H_2}$, the Dirac mass matrices are simply given by
\begin{equation}
    M^e_{ij} ~=~ \frac{v_{H_2}}{\sqrt{2}}\, c^e_{ij}\, \epsilon^{n^e_{ij}},
    \hspace{0.75cm}
    M^\nu_{ij} ~=~ \frac{v_{H_2}}{\sqrt{2}}\, c^\nu_{ij}\, \epsilon^{n^\nu_{ij}}.
\end{equation}
At leading order, the charged lepton masses are
\begin{eqnarray}
    \frac{m_e}{m_\tau}=\frac{(c^e_{12}-c^e_{23})^2}{c^{e^2}_{23}-1}\,\epsilon^4,
    \hspace{1.0cm}
    \frac{m_\mu}{m_\tau}=(1-c^{e^2}_{23})\,\epsilon^2.
\end{eqnarray}
Taking $\epsilon=0.1$, the following matrix of $c^e_{ij}$ coefficients reproduce the correct hierarchy between generations:
\begin{equation}
    c^e_{ij} ~=~ \begin{pmatrix}
        1.0  & ~1.6 & ~1.0  \\
        1.6  & ~1.0 & -2.7 \\
        1.0  & -2.7 & ~1.0 \\
        \end{pmatrix}\,.
\end{equation}
Since the tau mass is not suppressed by any additional factor, we expect $v_{H_2} = \epsilon^2\, v_{\rm EW}$, with $v_{\rm EW}\simeq 246\, {\rm GeV}$.
For this hierarchical scenario, the mixing pattern is
\bea
    \left|U^e_L\right|_{ij}=\left|U^e_R\right|_{ij} & \approx & \delta_{ij} + \epsilon^{n^e_{ij}}/\epsilon^{n^e_{jj}}\quad {\rm with~} i\leq j\,.
\eea
Then, the $e-\mu$ mixing is $\mathcal{O}(\epsilon)\sim 0.1$.
The masses of the active neutrinos are produced through the usual type-I seesaw mechanism.
Notice that, in this kind of formulations, the PMNS matrix can always be generated by a proper structure of the $M_R$-matrix \cite{Masiero:2002jn}.\\

After the breaking of the flavour symmetry, the flavon field can be parametrised as
\beq
    \phi ~=~ \frac{1}{\sqrt{2}}\, \left(v_\phi \,+\, s \right)\, e^{i\, a/v_\phi},
\eeq
with $s(x)$ a CP-even scalar and $a(x)$ the ALP.
If all the interactions respect the $\Ug\times Z_2$ symmetry, after the spontaneous breaking,  $a(x)$ should be the massless Nambu-Goldstone boson (NGB).
In our model, we included a soft-breaking term, $m^2 H_1 \cdot H_2$ to give a mass to it.
Alternatively, a hidden strong sector coupling to the ALP can be assumed. 
In the following, we treat the ALP mass as a free parameter and, as preferred by the Xenon1T data, it should be around the keV scale.\\

The interaction between the pseudo Nambu-Goldstone boson (pNGB) and the charged leptons, in the mass basis, is:
\beq \label{eq:Lae}
    -{\cal L}_{ae} ~=~ i\frac{\partial_\mu a}{2f_a}\: \overline{e}_i\, \gamma^\mu \left(V^e_{ij} + \gamma^5 A^e_{ij}\right) e_j,
\eeq
where $f_a \sim {\cal O}(v_\phi)$.
The axial and vector couplings in eq.\eqref{eq:Lae} are defined as\footnote{For $i=j$, we can always redefine the fields to have $V^e_{ii}=0$ \cite{Calibbi:2020jvd}}:
\bea
    V^e_{ij} & = & \frac{1}{2} \left(U^{e\,\dagger}_R x_R U^e_R \;+\; U^{e\,\dagger}_L x_L U^e_L\right), \label{eq:Ve} \\
    A^e_{ij} & = & \frac{1}{2} \left(U^{e\,\dagger}_R x_R U^e_R \;-\; U^{e\,\dagger}_L x_L U^e_L\right), \label{eq:Ae}
\eea
with $x_L$ and $x_R$ the diagonal $3\times 3$ matrices whose elements are the charged-lepton $\Ug$ charges and $U_L^e,\, U_R^e$ the unitary transformations that diagonalise the mass matrices\footnote{In our convention: $U^{e\,\dagger}_L\, M_e\, U^e_R={\rm Diag}(m_e,m_\mu,m_\tau)$.}.
In general, eqs.\eqref{eq:Ve} and \eqref{eq:Ae} induce FV effects which are subject to constraints from different experiments, as it is discussed in section \ref{sec:Constraints}.

\subsection{Model II: general mixing}
\label{sec:ModelII}
In model II, we generalize the previous structure to allow for arbitrary leptonic mixings.
To do this, we consider the $\Ug$ global symmetry as only part of a larger flavour symmetry, ${\cal F}$, that will determine the Yukawa structure with the observed hierarchy among generations in the lepton sector.
In this way, the $\Ug$ symmetry remains flavour dependent, but masses and mixings are not fixed by the $\Ug$ charges.\\
As an example, we use the same $\Ug\times Z_2$ charges as in Model I although now we can take $v_\phi/ \Lambda \simeq {\cal O}(1)$. 
The scalar potential and Yukawa terms remain as in eqs. \eqref{eq:VH} and \eqref{eq:LY} but, in this case, we highlight that the coefficients $c^e_{ij}$ and $c^\nu_{ij}$ are NOT forced to be ${\cal O}(1)$.\\
Adjusting them, different mixing patterns can be obtained.
In particular, we are interested in the case of large PMNS-like mixing for charged leptons.
As a typical benchmark model, we assume that the breaking of the symmetry $\cal F$ produces Yukawa couplings with PMNS-like mixing in the left- and right-handed sector. 
The couplings with the ALP are determined by eqs.\eqref{eq:Ve} and \eqref{eq:Ae}, but now
\bea
    V^e_{ij},\, A^e_{ij} & = & \frac{1}{2}\: U^{e\,\dagger}_{\rm PMNS}\, \Big(x_R \;\pm\; x_L\Big)\, U^e_{\rm PMNS}.
    \label{eq:VAeII}
\eea
Then, for example, we can deduce the size of the axial $12$-coupling to be as large as $A^e_{12}\simeq 0.56$.
Note that $V^e_{ij}=0$ by construction since $x_L=-x_R$.

\section{Constraints from LFV and astrophysics}
\label{sec:Constraints}
Non-universal charges of the charged leptons under the $\Ug$ global symmetry, together with non-trivial rotations to the mass basis, imply FV interactions between the ALP and these fermions.
The absence of the anomalous coupling between the ALP and photons at tree-level makes the search for ALPs by charged-lepton flavour-violating (CLFV) processes specially relevant.
Detailed discussions about the phenomenology of {\it flavourful} ALPs can be found in \cite{Bjorkeroth:2018dzu, Calibbi:2020jvd}.\\
Table \ref{tab:LFV} collects the experimental present limits and projected sensitivities for the search of ALPs through the detection of the FV process $\ell_i\to\ell_j\, a$.
For an ALP mass around the keV, the branching ratio for the FV transition $\ell_i \to \ell_j a$ is given by:
\begin{equation}
    \Br{\ell_i\to\ell_j a} \,=\, \frac{m_{\ell_i}^3}{16\pi\Gamma(\ell_j)}\frac{\left|C_{ij}^e\right|^2}{4\, f_a^2} \left( 1-\frac{m_a^2}{\ell_i^2}\right)^2.
\end{equation}
with $\left|C^e_{ij}\right|^2 = \left|V^e_{ij}\right|^2+\left|A^e_{ij}\right|^2$.
For a given model, where the interaction between the ALP and charged leptons is fixed, the bounds in Table \ref{tab:LFV} on the $\ell_i\to\ell_j a$ transitions can be translated into bounds on $f_a$.
Although all of them have been inspected, the strongest limits come from $\mu\to\e a$.\\
\begin{table}[h!]
	\centering
	{\renewcommand{\arraystretch}{1.25}
	\resizebox{0.45\textwidth}{!}{
	\begin{tabular}{l c l}
    \hline
    \bf Lepton decay & \bf \quad BR limit \quad\qquad & \bf Experiment \\[2.5pt]
    \hline
    \multirow{4}{*}{$\Br{\mu \to e\, a}$}
    & $< 2.6\cdot 10^{-6}$ & \texttt{Jodidio et al.} \cite{Jodidio:1986mz} \\[2.5pt]
    & $< 2.1\cdot 10^{-5}$ & \texttt{TWIST} \cite{Bayes:2014lxz} \\[2.5pt]
    & $< 1.3\cdot 10^{-7}$ & MEGII-fwd \cite{Calibbi:2020jvd}$^*$ \\[2.5pt]
    & $< 7.3\cdot 10^{-8}$ & \texttt{Mu3e} \cite{Perrevoort:2018okj}$^*$ \\[2.5pt]
    $\Br{\mu \to e\, a\, \gamma}$ & $< 1.1 \cdot 10^{-9}$ & \texttt{Crystal Box} \cite{Bolton:1988af} \\[2.5pt]
    \hline
    \multirow{2}{*}{$\Br{\tau \to e\, a}$} & $< 2.7\cdot 10^{-3}$ & \texttt{ARGUS} \cite{Albrecht:1995ht} \\[2.5pt]
    & $< 8.4\cdot 10^{-6}$ & \texttt{Belle-II} \\[2.5pt]
    \hline
    \multirow{2}{*}{$\Br{\tau \to \mu\, a}$} & $< 4.5\cdot 10^{-3}$ & \texttt{ARGUS} \cite{Albrecht:1995ht} \\[2.5pt]
    & $< 1.6\cdot 10^{-5}$ & \texttt{Belle-II} \\[2.5pt]
    \hline
    \end{tabular}}}
\captionsetup{width=0.45\textwidth,labelsep=none}
\caption{\label{tab:LFV} \small
        .- Limits over the axion decay constant from lepton decays.
        The $^*$ signals future bounds.
        \texttt{Belle-II} limits are derived from the simulated result at \texttt{Belle} \cite{Griessinger:2017rpx} by rescaling the luminosity \cite{Calibbi:2020jvd}.}
\end{table}

Regarding astrophysics bounds, interesting limits can be derived from stellar evolution.
In particular, the cooling of white dwarfs \cite{Bertolami:2014wua} (WD) and red giants \cite{Raffelt:1994ry, Viaux:2013lha} (RG) impose strong constraints over the ALP interactions to matter and radiation.
For massless ALP, the limits at $95\%$ CL are
\bea
    f_a & \gtrsim & 2.3\times 10^9\, \left|C^e_{11}\right|\, \GeV \,,\\
    f_a & \gtrsim & 1.2\times 10^9\, \left|C^e_{11}\right|\, \GeV\,.
\eea
For ALP masses above $1\, \keV$, the cooling rate is Boltzmann-suppressed 
and the limits above should be rescaled by the factor $\sqrt{\xi(m_a,T)/\xi(0,T)}$, where \cite{Viaux:2013lha}
\beq
    \xi(m_a,T) ~=~ \frac{1}{2\pi^2}\int^{\infty}_{m_a} \frac{E^2\sqrt{E^2-m_a^2}}{e^{E/T}-1}.
\eeq
Finally, we comment on an additional bound from the Big Bang Nucleosynthesis (BBN).
As discussed in \cite{Ghosh:2020vti}, by constraining the effective number of relativistic neutrino species during BBN, non-negligible limits can be set over the lepton-ALP coupling that compete with those from stellar evolution.
When $m_a\sim [20,\, 1000]\, \keV$, the BBN bound is indeed the strongest one.
In our scenario, however, with a few keV ALP mass, stellar evolution remains the most important phenomena.

\section{RESULTS}
\label{sec:results}
In \cite{Takahashi:2020bpq}, the authors conclude that an ALP satisfying
\begin{equation} \label{eq:resulXen1T}
    A^e_{11}\simeq 10^{-13}\frac{f_a}{m_e},\, \hspace{0.5cm} {\rm for}\; m_a\in [2,3]\,\keV,
\end{equation}
can reproduce the Xenon1T signal, together with some reported anomalies in stellar cooling \cite{Viaux:2013lha, Bertolami:2014wua}.
In the same work, it is argued that such possibility can be realised in the context of anomaly-free DM ALPs, provided that the ALP constitutes only a $7\%$ of the total DM abundance. 
The discussion is however restricted to astrophysical and cosmological constraints and flavour observables are not discussed.
Here, we aim to highlight the role of flavour observables to (dis)prove this kind of models.\\
\begin{figure}[h!]
    \centering
        \includegraphics[width=0.48\textwidth]{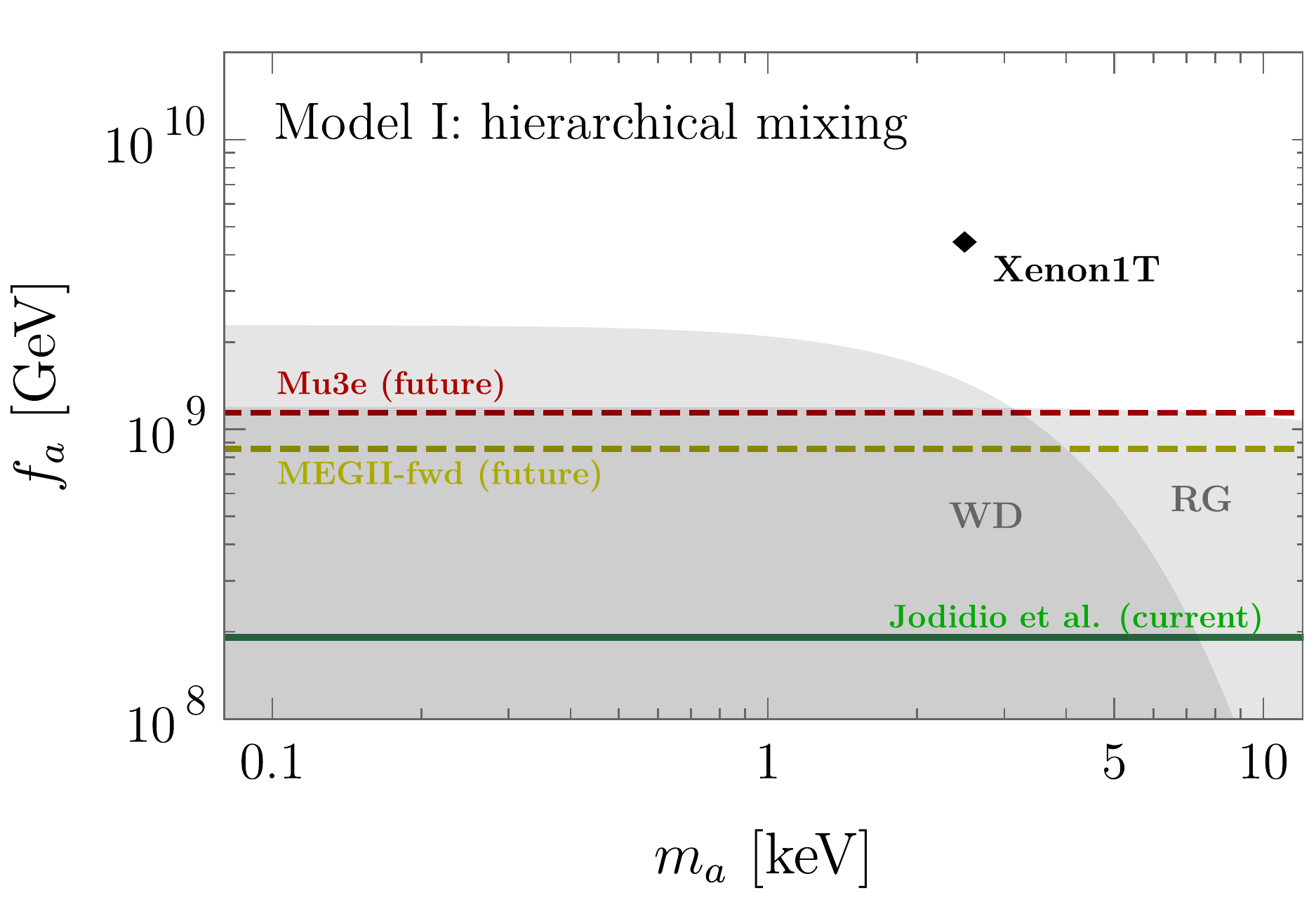}
    \captionsetup{width=0.5\textwidth, 
    }
    \caption{Results for Model I with hierarchical Yukawa couplings generated à la Froggatt-Nielsen.}
    \label{fig:ModI_results}
\end{figure}
\begin{figure}[h!]
    \centering
        \includegraphics[width=0.48\textwidth]{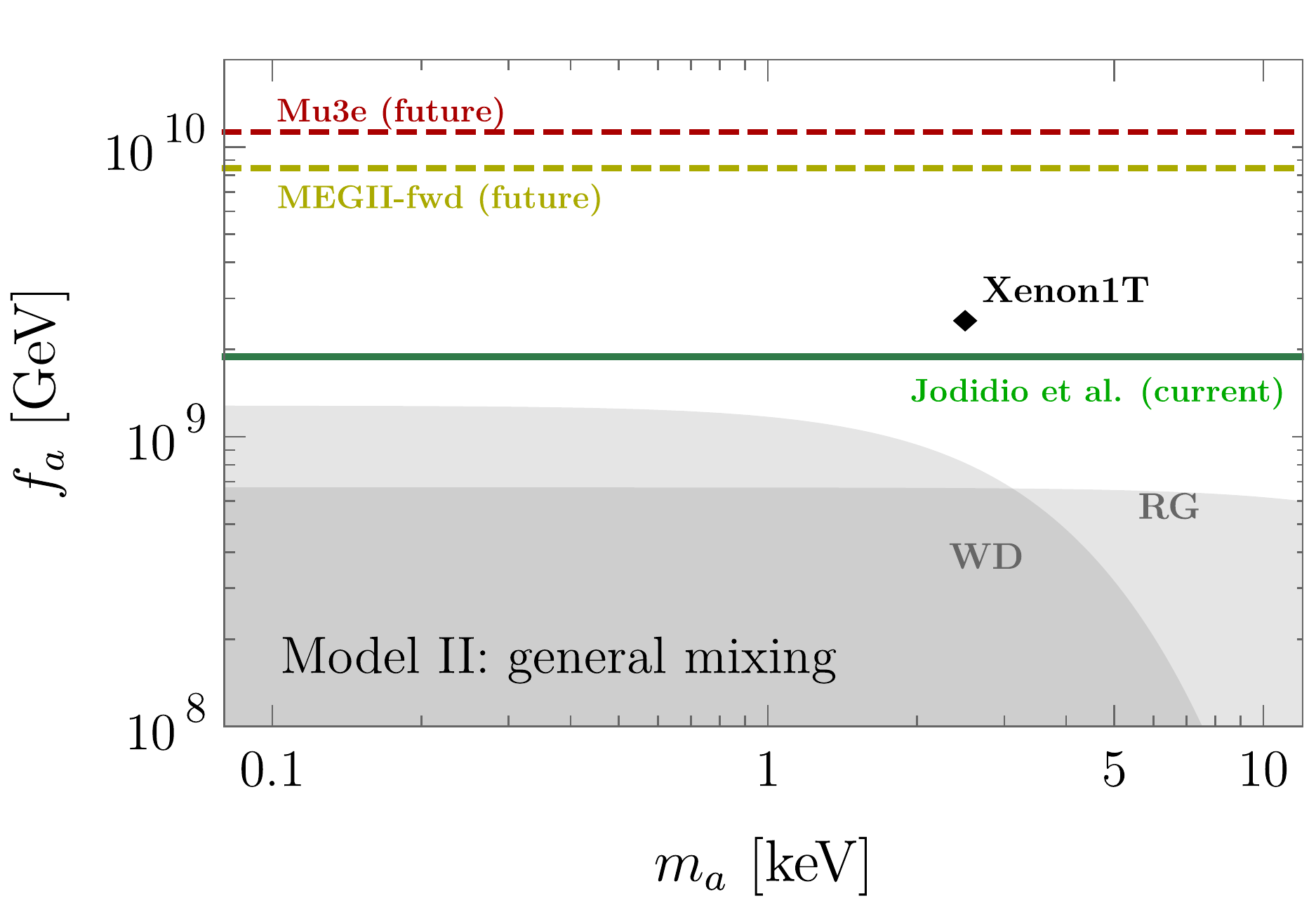}
    \captionsetup{width=0.5\textwidth
    }
    \caption{Results for Model II with general Yukawa matrices and mixing.}
    \label{fig:ModII_results}
\end{figure}

Figures \ref{fig:ModI_results} and \ref{fig:ModII_results} show the Xenon1T favoured prediction for $f_a$, based on the result in eq.\eqref{eq:resulXen1T} (black diamond).
Similarly, current and expected sensitivity from \texttt{Jodidio et al.} \cite{Jodidio:1986mz} (green continuous line) and \texttt{Mu3e} \cite{Perrevoort:2018okj} (red dashed line) in dedicated searches for $\mu\to\e\gamma$ are displayed as a function of $m_a$.
We also show the projection of the proposal by Calibbi et al. \cite{Calibbi:2020jvd}, MEGII-fwd (yellow dashed line), for \texttt{MEGII} \cite{Baldini:2018nnn} to improve the detection of the process of interest, $\mu\to e\, a$.
Finally, limits due to white dwarfs and red giants (gray shaded regions) also impose relevant bounds on our models \cite{Calibbi:2020jvd}.\\
From figure \ref{fig:ModI_results}, we notice that testing Model I (small mixing) with LFV observables remains quite challenging, even for future sensitivities.
On the other hand, scenarios with larger mixing effects in the charged-lepton sector provide better prospects.
For Model II, in figure \ref{fig:ModII_results}, we observe that while current limits are not sufficient to constrain the model, more stringent bounds coming from \texttt{Mu3e} or the implementation of MEGII-fwd are enough to probe this formulation.
One may then conclude that LFV can clearly complement astrophysics searches and, in some cases, go beyond them.
Flavoured ALP models provide a rich phenomenology to be investigated with present and future data.\\

\begin{figure}
    \centering
        \includegraphics[width=0.235\textwidth]{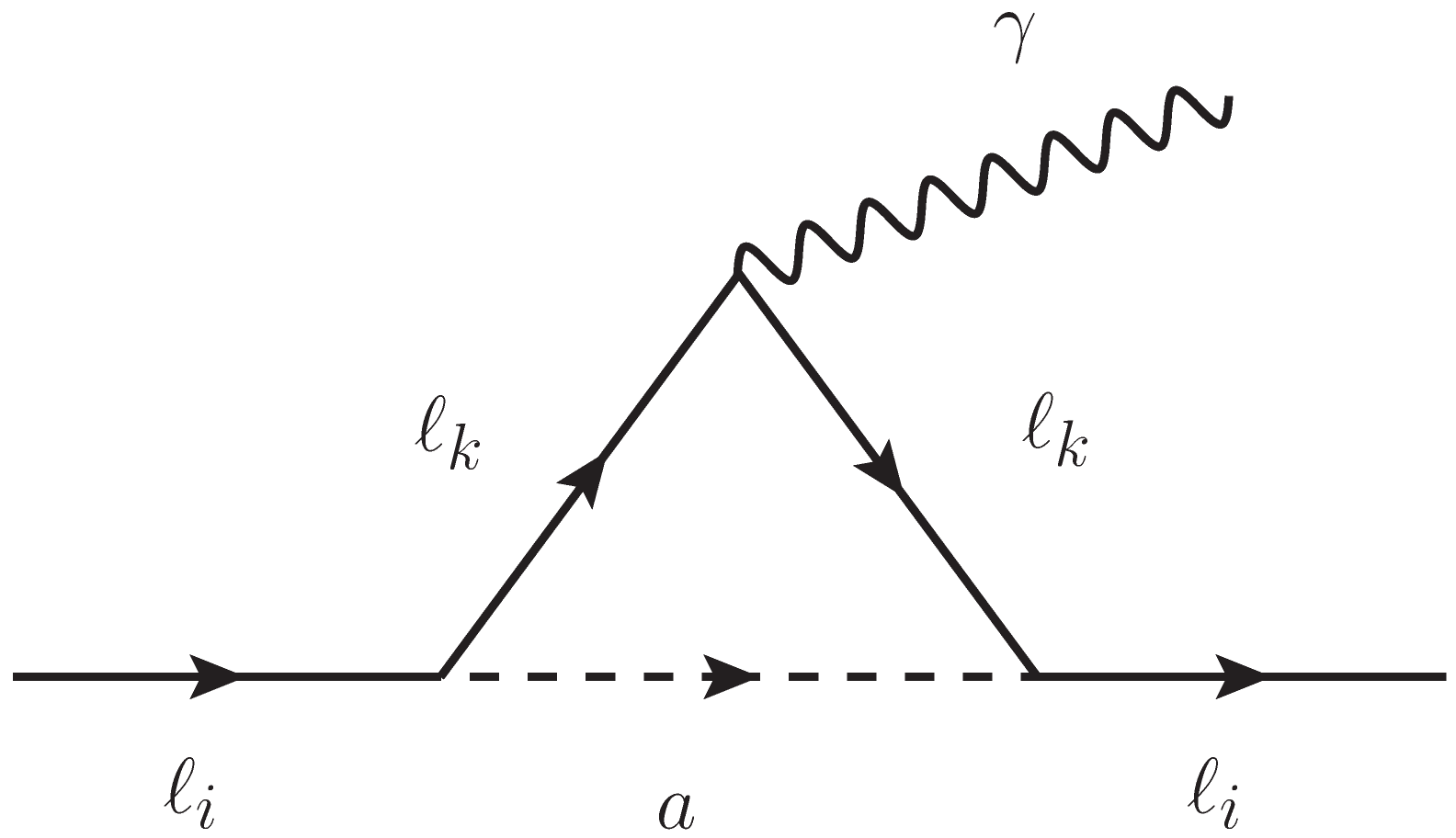}
        \includegraphics[width=0.235\textwidth]{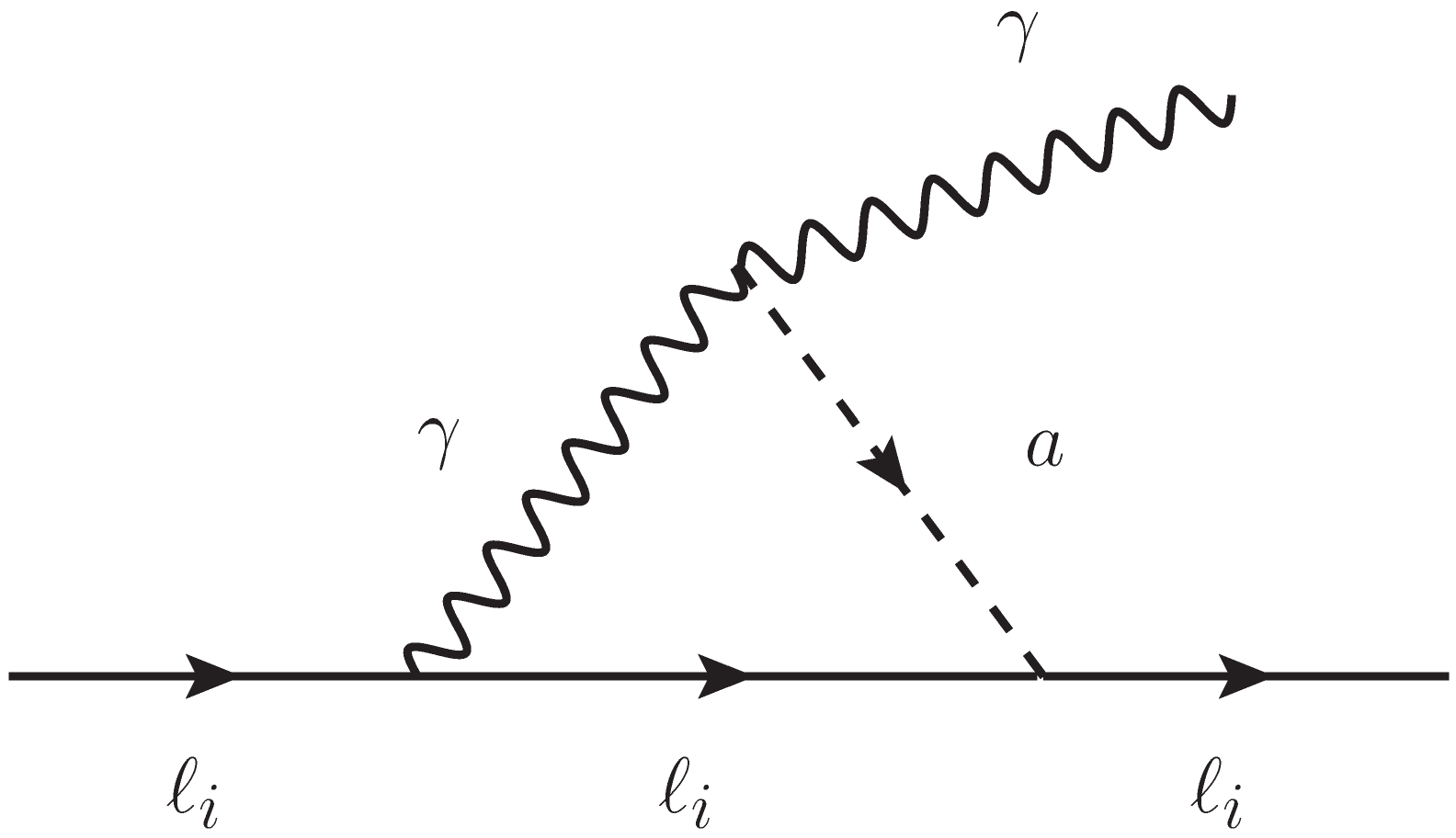}
    \captionsetup{width=0.5\textwidth
    }
    \caption{Diagrams contributing to $(g-2)_\ell$. 
    }
    \label{fig:g-2}
\end{figure}

A final remark about the ALP solution to $(g-2)_e$ and $(g-2)_\mu$ can be made at this point.
Two type of contributions have been discussed in the literature to explain the observed discrepancies, involving flavour-conserving and flavour-violating interactions between an ALP and charged leptons \cite{Chang:2000ii, Marciano:2016yhf, Bauer:2017ris, Bauer:2019gfk, Cornella:2019uxs, Endo:2020mev}.
The size of these effects has been evaluated for the models discussed here.
The relevant processes are those in figure \ref{fig:g-2}, which correspond to LFC (left, {\it i=k}), LFV (left, {\it i$\neq$k}) and (LFC) Barr-Zee type transitions (right) \cite{Cornella:2019uxs}.
The latter is highly suppressed in our framework due to the absence of tree-level couplings between the ALP and photons, so we focus on the first diagram.\\
In Model I, both $\Delta a_\mu$ and $\Delta a_e$ can be reproduced simultaneously at $1\sigma$ and $2\sigma$ respectively for $m_a=2\, \keV$ if $f_a\sim 70-90\, \GeV$.
For electrons, we find that the LFC diagram is dominant and negative so that 
$\Delta a_e\simeq -m_e^2/(16\pi^2)\left|A^e_{11}\right|^2/f_a^2$.
Instead, muons receive the main contribution from the LFV transition involving tau leptons in the loop, which turns out to be positive:
$\Delta a_\mu\simeq m_\mu m_\tau/(64\pi^2)\big(\left|V^e_{23}\right|^2-\left|A^e_{23}\right|^2\big)/f_a^2$.
In Model II, the anomalous magnetic moments are mainly determined by the LFV transitions, $\Delta a_\ell\simeq m_\ell m_j/(64\pi^2)\big(|V^e_{\ell j}|^2-|A^e_{\ell j}|^2\big)/f_a^2$, which are always negative since $V^e_{ij}=0$.
That is in agreement with the electron measurement, which can be explained if $f_a\sim220\, \GeV$ for $m_a=2\, \keV$, but in conflict with $\Delta a_\mu$, which cannot be generated with the correct sign.\\
In any case, notice that the resulting values for $f_a$ are too low to be consistent with the LFV limits in table \ref{tab:LFV}, see figures \ref{fig:ModI_results} and \ref{fig:ModII_results}.
Moreover, they induce unacceptable large branching fractions for several leptonic decays of the type $\ell_j\to \ell_i \gamma$ and $\ell_j \to 3\ell_i$.
We, therefore, conclude that our models cannot provide a successful explanation to the observed anomalies.
\section{CONCLUSIONS}
\label{sec:conclusions}
In this paper, we considered the LFV effects from a keV scale flavour-dependent ALP which is motivated by recent Xenon1T results. We find that, for a general mixing in the lepton sector, the leptonic flavour changing experiments could confirm or exclude the possibility of explaining the Xenon1T result  by an ALP, while being consistent with all phenomenological and astrophysical constraints. On the other hand, if the leptonic mixing originating from the Froggatt-Nielsen symmetry are small, CKM-like, the measurement of their LFV effects would constitute a challenge for future experiments.
\begin{acknowledgments}
The authors thank Arcadi Santamaria for useful discussions.
AM acknowledges support from La-Caixa-Severo Ochoa scholarship.
AM and OV are supported by Spanish and European funds under MICIU Grant FPA2017-84543-P.
OV acknowledges partial support from the “Generalitat Valenciana” grant PROMETEO2017-033.
JMY acknowledges funding from the National Natural Science Foundation of China (NNSFC) under grant Nos.11675242, 11821505, and 11851303, from Peng-Huan-Wu Theoretical Physics Innovation Center (11947302), from the CAS Center for Excellence in Particle Physics (CCEPP), from the CAS Key Research Program of Frontier Sciences and from a Key R$\&$D Program of Ministry of Science and Technology under number 2017YFA0402204. CH acknowledges support from the Sun Yat-Sen University Science Foundation.
\end{acknowledgments}



\end{document}